\documentclass[aps,pra,twocolumn,superscriptaddress,eqsecnum]{revtex4}
\usepackage{graphicx}
\usepackage{bm}
\usepackage{subfig}
\usepackage{amsmath,amssymb,amsthm,dsfont,bm}
\usepackage{color}
\usepackage{wasysym}
\usepackage{ulem}
\usepackage{cancel}
\usepackage{verbatim}
\usepackage{soul}

\usepackage[applemac]{inputenc}
\usepackage[dvipsnames]{xcolor}
\usepackage[colorlinks=true,urlcolor=blue,citecolor=blue,linkcolor=blue]{hyperref}

\begin{document}

\title{Coherence and incoherence in quadrature basis}
\author{Laura Ares}
    \affiliation{Theoretical Quantum Science, Institute for Photonic Quantum Systems, Paderborn University, Warburger Stra\ss{}e 100, 33098 Paderborn, Germany}
    
\author{Alfredo Luis}
    \affiliation{Departamento de \'Optica, Facultad de Ciencias
F\'{\i}sicas, Universidad Complutense, 28040 Madrid, Spain}

\date{\today}

\begin{abstract}
How to manage coherence as a continuous variable quantum resource is still an open question. We face this situation from the very definition of incoherent states in quadrature basis. We apply several measures of coherence for some physical states of light relative to a quadrature basis. We examine the action on the coherence of several transformations such as beam splittings and squeezing.
\end{abstract}

\maketitle

\section{Introduction}

Quantum coherence lies at the heart of quantum mechanics, rooted in its wave-like nature. It is the cause of the technological advantages offered by quantum effects in diverse areas such as quantum computing, communication, or resolution. By shedding light on the many sides of coherence, we can improve our understanding of quantum theory and its applications.

\bigskip

Nowadays, one of the most common and successful approaches to quantum coherence is the resource theory of coherence. By construction, this framework is mainly focused on finite-dimensional spaces and discrete variable systems. However, continuous variables have a great impact as resources for quantum applications \cite{SLB05}. Therefore, they are demanding a great deal of research given their complexity \cite{ZSLF16,GF23}. The switch from discrete to continuous variables is usually focused on Gaussian states \cite{GA18}. As a particular example, we can find the resource theory of squeezing and nonclassicality \cite{BY18,MI16}.  As an example of different approaches to continuous variable resources, we can find scenarios based on robustness \cite{BR21,LL21,HKPU21}.Therefore, this subject is plenty of subtle details that certainly may help to a deeper, global knowledge of the concept of coherence, and the quantum-optical field quadratures offer a distinguished arena to examine these points. 

\bigskip

Quadratures play a distinguished role among quantum-optical observables. They have exactly the same meaning in the quantum and classical realms, with clear mechanical counterparts in the form of position and linear momentum observables. These observables are easily measured in all regimens, including low and large field intensities. Moreover, they are key observables regarding metrology, detection and state reconstruction. This bring us  to the study of coherence in the quadrature basis, with special attention to transformation properties which are basic to coherence assessment. 

\bigskip

In this work we approach the concept of coherence in quadrature trying to mimic to the maximum the main points of discrete variable scenarios. To this end we utilize the concept of incoherent state for unnormalizable basis introduced in Ref. \cite{AL22} as an extension of the discrete variable and finite dimensional scheme. These incoherent states arise as the key difference with previous approaches, as can be seen from the analysis of incoherent or free  operations: those transforming incoherent states into incoherent states. As a specific question, we study the action of incoherent operations on partially coherent states. We study the performance of this coherence by means of some significant examples and compare the implications of using different quantifiers. There we find the possibility that incoherent operations may increase the coherence of partially coherent states. The question of creating resources by means of free operations has been already addressed \cite{LGT17,LZ20}, revealing the existence of coherence monotones under strictly incoherent operations that can increase under incoherent operations.

\bigskip

\section{Coherence for quadratures}

The system under study is a single-mode field with complex-amplitude operator $a$ and quadratures, $X$ and $Y$, defined as the real and imaginary parts of the complex amplitude, $a= X + iY$. We may as well consider them {\it position} or {\it linear momentum} of a particle or an harmonic oscillator. 

Moreover, since the distinction between quadratures is merely a question of a choice of Cartesian axes in the complex plane, this is a phase, we may as well consider a generic quadrature $X_\theta$ as 
\begin{equation}
    X_\theta = \frac{1}{2} \left ( a e^{-i \theta} +  a^\dagger e^{i \theta} \right ) .
\end{equation}

\bigskip
As coherence measure we consider an straightforward generalization of l1-norm of coherence in finite-dimensional spaces with numerable basis \cite{BCP14}: 
\begin{equation}
\label{mc}
    \mathcal{C} = \int_{-\infty}^\infty\int_{-\infty}^\infty dx dx^\prime | \langle x |\rho | x^\prime \rangle | ,
\end{equation}
where $|x\rangle$ are the eigenstates, not normalizable, of some quadrature $X_\theta$, say $X=X_{\theta =0}$. For pure states $\rho = | \psi \rangle \langle \psi |$ this can be expressed in terms of the  quadrature {\it wave-function}, $\psi, (x)$ as
\begin{equation}
    \mathcal{C} = \left ( \int_{-\infty}^\infty |\psi (x) | \right )^2 ,  \quad  \psi (x) = \langle x |\psi \rangle .
\end{equation}

In Ref. \cite{AL22}, we have shown that this generalization has desirable properties regarding geometrical considerations as distance to the incoherent states. For the sake of comparison, we similarly present another measure in terms of relative entropy. We may recall here previous measures of coherence in the quadrature basis such as the coherence quadrature scale \cite{HB20,GTH23}. 

\bigskip

\section{Incoherent states and relative entropy}

Incoherent states are diagonal in the reference basis, formed in this case by the unnormalizable eigenstates  $|x\rangle$ of the quadrature $X$. However, there is no physical state admitting the decomposition
\begin{equation}
\label{pps}
    \rho_{\rm in} =  \int dx P(x)  |x \rangle \langle x | ,
\end{equation}
due to the lack of finite $\mathrm{tr} \rho_{\rm in}$. Thus, we should address this subject from the very definition of the free states.

\bigskip

\subsection{Incoherent states}

As shown in Ref. \cite{AL22}, there are at least two relevant families of physical states that tend to be incoherent in the appropriate limit, allowing an approach as close as possible to the most standard formulation of coherence. On the one hand, Gaussian squeezed states in the limit of infinite squeezing, $\Delta X \rightarrow 0$. On the other hand, thermal states in the limit of an arbitrarily large mean number of photons, $\overline{n} \rightarrow \infty$. In both cases, the quantum coherence of these states tends to be null, $\mathcal{C} \rightarrow 0$. 

\bigskip

Finally, we may add a family of no so physical states $ |\chi_{j,\sigma} \rangle$, that maybe nevertheless useful for the analysis. They emerge from a discretization of the $x$ axis so their wave-function $ \langle x |\chi_{j,\sigma} \rangle$ is the characteristic function of the $x$ interval centered at point $x_j = j \sigma$ with width $\sigma$, this is
\begin{equation}
\label{chi}
\langle x |\chi_{j,\sigma} \rangle = \left \{ \begin{array}{l}  \frac{1}{\sqrt{\sigma}} \;  \mathrm{for} \; x \in (j \sigma- \sigma/2, j \sigma + \sigma/2 ),  \cr  \cr 0 \;\mathrm{for} \; x \notin (j \sigma - \sigma/2, j \sigma + \sigma/2 ) ,  \end{array} \right .  
\end{equation}
and in this case the proper limit is $\sigma \rightarrow 0$. The property that may render this family useful in front of the other two is orthogonality 
\begin{equation}
\label{orto}
    \langle \chi_{j, \sigma} | \chi_{k, \sigma} \rangle = \delta_{j,k} .
\end{equation}

\color{black}
\bigskip

We cannot forget that these limits imply arbitrary high energy. This connection incoherence-energy is rather interesting. In the  classical-optics domain there is no relation between coherence and energy, while in the quantum-optics realm there is the increase of coherence with energy \cite{AL22}.  This feature of infinite-limit energy has been solved in different scenarios by defining the free states as those with the minimum amount of resource, even if it is not cero \cite{GA18,HB20}.

\bigskip

With this in mind, we try to overcome the unnormalization of the elements of the basis in the definition of the incoherent state. To this end, we replace $|x\rangle$ in Eq. (\ref{pps}) by some normalizable states, for example the displaced-squeezed states $|\xi_{x,\sigma} \rangle$ with quadrature-coordinate wave function
\begin{equation}
\label{wf}
    \langle x^\prime|\xi_{\bar{x},\sigma} \rangle= \frac{1}{(2 \pi \sigma^2 )^{1/4}} \exp \left [-\frac{(x^\prime-\bar{x})^2}{4 \sigma^2} \right ],
\end{equation}
where $\bar{x} = \langle  \xi_{\bar{x},\sigma} | X | \xi_{\bar{x},\sigma} \rangle$ and $\sigma = \Delta X$ are the mean value and standard deviation of quadrature $X$ respectively.

Then, we can define a unit-trace states, $\rho_{\rm in}$, as
\begin{equation}
\label{rdxc}
\rho_{\rm in} =  \int_{-\infty}^\infty dx  P(x) |\xi_{x,\sigma}\rangle \langle \xi_{x,\sigma} | ,
\end{equation} 
for any probability distribution $P(x)$. Taking into account that the wave function in Eq. (\ref{wf}) is real and nonnegative, we get 
\begin{equation}
 \mathcal{C} (\rho_{\rm in}) =  \int_{-\infty}^\infty dx P(x)  \mathcal{C} (|\xi_{x,\sigma}\rangle),
\end{equation}
and given that
\begin{equation}
    \mathcal{C} (|\xi_{x,\sigma}\rangle) =  2 \sqrt{2\pi} \sigma ,
\end{equation} 
is independent of $x$, we finally get
\begin{equation}
\label{Cin}
 \mathcal{C} (\rho_{\rm in}) =  2 \sqrt{2\pi} \sigma .
\end{equation}

We may consider incoherent states those satisfying some minimum nonzero $\mathcal{C}$ under some restrictions \cite{GA18,HB20}. Instead, we  consider that incoherence manifests in the limit $\sigma \rightarrow 0$, so that $\mathcal{C} (\rho_{\rm in}) \rightarrow 0$. It is worth noting that there is no contribution of $P(x)$ to the coherence.

\bigskip

We may use this idea to approach the diagonal part of any density matrix $\rho$ in the basis $X$ as 
\begin{equation}
\label{rd}
\rho_d =  \int_{-\infty}^\infty dx  \langle x | \rho | x \rangle  |\xi_{x,\sigma}\rangle \langle \xi_{x,\sigma} |,
\end{equation} 
as the incoherent state $\rho_{\rm in}$ closer to $\rho$ , as it can be checked for example via the Hilbert--Schmidt distance, in the limit $\sigma \rightarrow 0$ 
\begin{equation}
    \mathrm{tr} \left [ \left ( \rho-\rho_{\rm in} \right )^2 \right ] .
\end{equation}
\bigskip

Similarly, we may utilize the incoherent states introduced in Eq. (\ref{chi}), so that the  incoherent part of $\rho$ becomes
\begin{equation}
\label{rd2}
\rho_{ d} =  \sum_{j=-\infty}^\infty\langle\chi_{j,\sigma} | \rho | \chi_{j,\sigma}\rangle  |\chi_{j,\sigma}\rangle \langle \chi_{j,\sigma} | ,
\end{equation}
where 
\begin{equation}
\langle\chi_{j,\sigma} | \rho | \chi_{j,\sigma}\rangle \simeq  \sigma p (j \sigma ) , \quad p(x) = \langle x| \rho | x \rangle  ,
\end{equation} 
where $| x \rangle$ are the unnormalized quadrature eigenstates. Then, in the limit $\sigma \rightarrow \infty$, the trace becomes 
\begin{equation}
\sum_{j=-\infty}^\infty\langle\chi_{j,\sigma} | \rho | \chi_{j,\sigma}\rangle  \rightarrow \int p(x) dx = 1 .
\end{equation} 

\subsection{Relative entropy}

\bigskip

Let us address the relative-entropy between $\rho$ and its incoherent part, $\rho_d$,
\begin{equation}
\label{re}
    S(\rho||\rho_d) = \mathrm{tr} \left ( \rho \ln \rho \right ) - \mathrm{tr} \left ( \rho \ln \rho_d \right ) .
\end{equation}
The first factor is rather universal, so we may focus on the second factor. In this regard, for the incoherent part $\rho_d$ we may consider either the expression (\ref{rd}) or (\ref{rd2}). The latter is specially adequate in this case since we can use the explicit orthogonality (\ref{orto}), so that 
\begin{equation}
   \mathrm{tr} \left ( \rho \ln \rho_d \right ) = \sum_{j=-\infty}^\infty \langle\chi_{j,\sigma} | \rho |\chi_{j,\sigma} \rangle \ln  \langle\chi_{j,\sigma} | \rho |\chi_{j,\sigma} \rangle  ,
\end{equation}
this is 
\begin{equation}
   \mathrm{tr} \left ( \rho \ln \rho_d \right ) \simeq  \sum_{j=-\infty}^\infty \sigma p(j \sigma) \ln  \left [ \sigma p(j \sigma)  \right ].
\end{equation}
We recall that $p(x)$ is the quadrature distribution in the state $\rho$, 
so this term goes as 
\begin{equation}
   \mathrm{tr} \left ( \rho \ln \rho_d \right ) \rightarrow  \ln \sigma + \int_{-\infty}^\infty dx p(x) \ln p(x) .
\end{equation}
The first factor $\ln \sigma$ diverges as $\sigma \rightarrow 0$, while the second factor is the usual form of Shannon entropy for a continuous distribution. At this point, the diverging $\ln \sigma$ may be safely ignored, since it is a known instrumental factor revealing that entropy in the continuous case is not exactly the limit of the discrete case \cite{CT06}. In any case, this factor  does not affect the comparison of the relative entropy between different states. Nevertheless, this divergence may obscure the idea of incoherent states since the splitting of the $\sigma$ factor may affect the limit $\sigma \rightarrow 0$.

\color{black}

\section{Coherence of some meaningful state families}

Let us apply this formalism to some relevant field states in quantum optics. 

\bigskip

\subsection{Gaussian pure states} 

We have already addressed this case regarding incoherent states. This includes coherent and squeezed states, assuming for simplicity that they are minimum uncertainty states, $\Delta X \Delta Y =1/4$, where $Y= X_{\theta=\pi/2}$,
\begin{equation}
\label{Gs}
   \psi (x) = \frac{1}{\left ( 2 \pi \Delta^2 X \right )^{1/4}} \exp \left [2 i \overline{Y} x - \frac{\left ( x-\overline{X} \right )^2}{4 \Delta^2 X} \right ] ,
\end{equation}
we have 
\begin{equation}
\label{CG}
       \mathcal{C} = 2 \sqrt{2\pi} \Delta X .
\end{equation}

\bigskip

As well illustrated by this Gaussian example, the fluctuations of the quadratures may play a relevant role according to the general intuition that coherence has to do with randomness. It is worth noting the duality relation between coherence in complementary quadratures, and the direct relation between $\mathcal{C}$ and quantum Fisher information under displacements generated by $X$.

\bigskip

Alternatively, the relative measure of coherence results
\begin{equation}
    S(\rho||\rho_d) = \frac{1}{2} \left [ 1+ \ln \left ( 2 \pi \Delta^2 X \right ) \right ] ,
\end{equation}
with the same behavior than $\mathcal{C}$. 

\color{black}

\bigskip

\subsection{Thermal states}

These are the mixed states 
\begin{equation}
    \rho = \frac{1}{1+\overline{n}} \sum_{n=0}^\infty \left ( \frac{\overline{n}}{1+\overline{n}} \right )^n |n \rangle \langle n |,
\end{equation}
where $| n \rangle$ are number states, and $\overline{n}$ the mean number of photons. Equivalently, 
\begin{equation}
\label{Ts}
    \rho =  \frac{1}{\pi \overline{n}} \int d^2 \alpha \exp \left ( - \frac{|\alpha |^2}{\overline{n}} \right ) |\alpha \rangle \langle \alpha |  ,
\end{equation}
where $| \alpha \rangle$ are coherent states, $\alpha = x+iy$, $d^2 \alpha = dx dy$. With this, and using Eq. (\ref{Gs}) for coherent states $\Delta X = 1/2$, we get  
\begin{equation}
 \langle x |\rho | x^\prime \rangle = \sqrt{\frac{2/\pi}{1+2\overline{n}}} \exp \left ( - \bm{x}^t M \bm{x} \right ) ,
\end{equation}
where $\bm{x}^t = (x, x^\prime)$, the superscript $t$ denotes transposition, and
\begin{equation}
    M = \frac{1}{1+2\overline{n}}\left ( \begin{array}{cc} 1 & 0 \\ 0 & 1 \end{array} \right ) + \frac{2 \overline{n} \left ( 1 + \overline{n} \right )}{1+2\overline{n}}  \left ( \begin{array}{cc} 1 & -1 \\ -1 & 1 \end{array} \right ) ,
\end{equation}
leading to
\begin{equation}
       \mathcal{C} = \sqrt{\frac{2\pi}{1 + 2 \overline{n}}} .
\end{equation}

\bigskip

Curiously, this coherence for thermal states is directly related to the quadrature uncertainty, since for any $\theta$
\begin{equation}
   \Delta X_\theta = \frac{\sqrt{1+2 \overline{n}}}{2} ,
\end{equation}
and then 
\begin{equation}
     \mathcal{C} =\frac{ \sqrt{2\pi}}{2 \Delta X_\theta} .
\end{equation}

\bigskip

Regarding relative coherence we get 
\begin{eqnarray}
\label{Sth}
     & S(\rho||\rho_d) = \ln \left [ \frac{1}{1+\overline{n}} \left ( \frac{\overline{n}}{1+\overline{n}} \right )^{\overline{n}} \right ] & \nonumber \\
     & + \frac{1}{2} \left \{ 1+ \ln \left [ \frac{\pi}{2} \left ( 1+2 \overline{n} \right ) \right ] \right \}  ,&
\end{eqnarray}
with the same behavior than $\mathcal{C}$, this is $S$ decreases for increasing $\overline{n}$. 

\color{black}

\bigskip

\subsection{Number states}

The number states have a $x$ wave-function
\begin{equation}
\psi (x)=\sqrt{\frac{2}{2^n n!\sqrt{2\pi}}} H_n\left(\sqrt{2} x \right) e^{ -x^2} .
\end{equation}

\bigskip

In this case, we have no analytic expression for $ \mathcal{C}$. In Fig. 1, it is represented a numerical evaluation of the quotient of the coherence for number states, $\mathcal{C} (n)$, versus the coherence for Gaussian states, $\mathcal{C} (\overline{n})$, with the same mean number of photons $n= \overline{n}$. This shows the Gaussian states are more efficient than number states regarding coherence for the same energy.

\begin{figure}
\includegraphics[width=8cm]{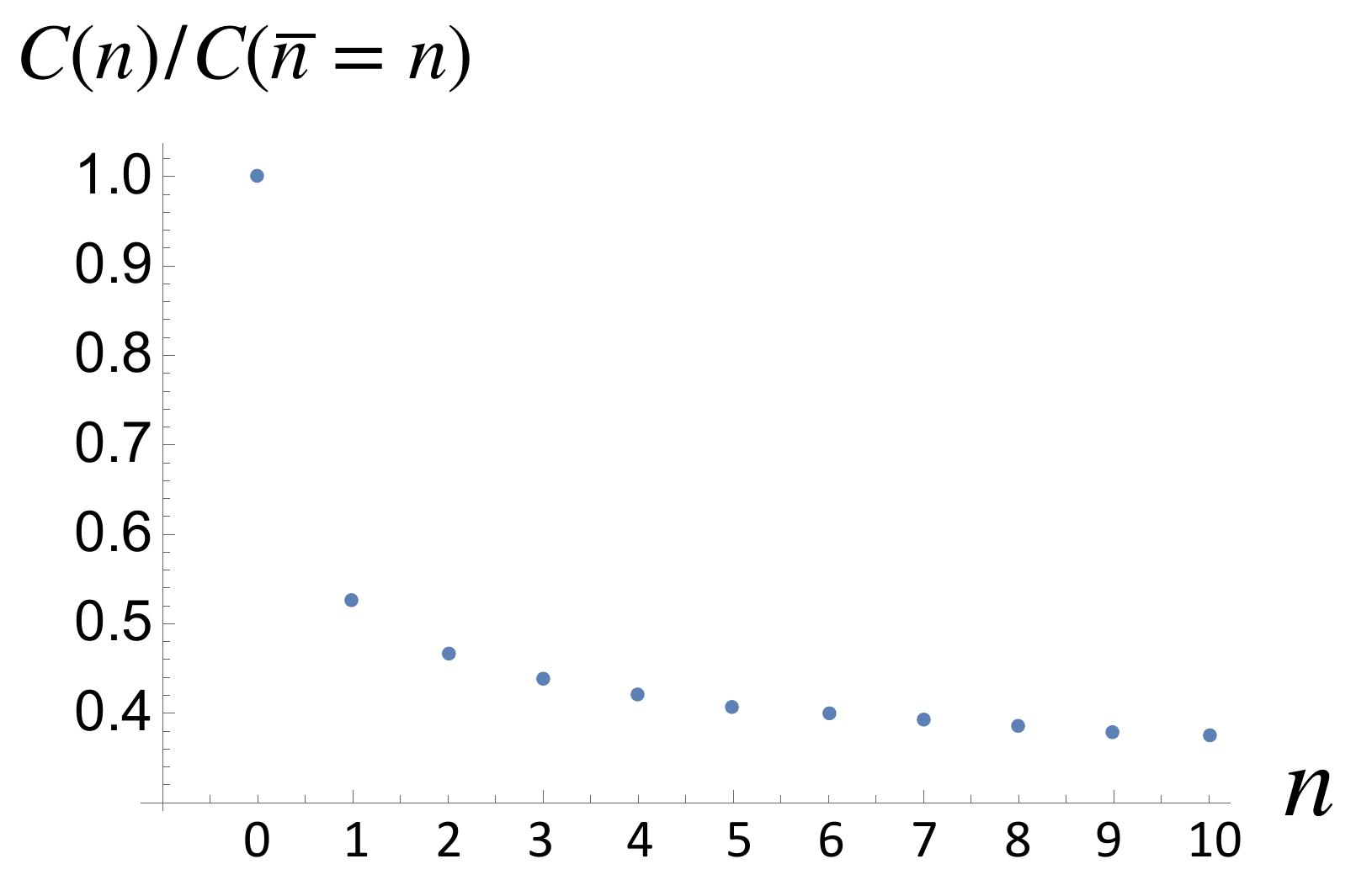}
\caption{Numerical evaluation of the quotient $\mathcal{C} (n)/\mathcal{C} (\overline{n})$ of the coherence for number states versus the coherence for Gaussian states with the same mean number of photons $n= \overline{n}$ as a function of $n$.}
\end{figure}{}

\section{Transformation of coherence}

In this section, we examine the behavior of this quadrature coherence under basic quantum-optical transformations. We focus on the measure (\ref{mc}), but a similar conclusion is obtained in terms of relative entropy.

\subsection{Free evolution}

Free evolution is generated by the number operator, $a^\dagger a$, as $U=\exp (- i \tau a^\dagger a)$ for a suitably scaled dimensional  time $\tau$. This transformation is a rotation in the complex-amplitude plane so that 
\begin{eqnarray}
 & X_{\tau} = U^\dagger X U = \cos \tau X + \sin \tau Y , & \nonumber \\
  & & \\
& U^\dagger Y U = - \sin \tau X + \cos \tau Y . & \nonumber
\end{eqnarray}

\bigskip

To illustrate this evolution we start considering the Gaussian case. The evolution of a Gaussian pure state is always a Gaussian pure state with different parameters $\overline{X}$, $\overline{Y}$, and $\Delta X$. Therefore,   after Eq. (\ref{CG}) we obtain
\begin{equation}
  \mathcal{C} (\tau) = 2 \sqrt{2\pi} \Delta X_\tau ,
\end{equation}
leading to, for the Gaussian states (\ref{Gs}) that lack initial correlations between quadratures $X$ and $Y$,
\begin{equation}
  \mathcal{C} (\tau) = 2 \sqrt{2\pi} \sqrt{\cos^2 \tau \Delta^2 X + \sin^2 \tau \Delta^2 Y}. 
\end{equation}
This is a periodic variation of coherence between the two extreme values, say
\begin{equation}
   2 \sqrt{2\pi} \Delta X, \qquad   2 \sqrt{2\pi} \Delta Y.
\end{equation}
Thus, free-field evolution would not be an incoherent transformation in this approach. Regarding thermal and number states there is no effect of evolution since they are stationary states.

\bigskip

\subsection{Displacements}

Displacements on the complex-amplitude plane are of the form of unitary transformations $D$ such that 
\begin{equation}
D^\dagger X D =  X + x_0 , \qquad D^\dagger Y D = Y + y_0 ,
\end{equation}
where $x_0$ and $y_0$ are real numbers. Displacements preserve the eigenstates of $X$ with a shift of the eigenvalue. Such a shift can be then absorbed with a change of variables in the $x, x^\prime$ integrals in Eq. (\ref{mc}) so that there is no effect of displacements on coherence for any state, and therefore these are incoherent operations. This agrees with postulate 0 in \cite{GA18}.

\bigskip

\subsection{Squeezing}

 Now we consider the unitary quadrature-squeezing transformation $T$,
\begin{equation}
\label{qst}
     T^\dagger X T = \lambda X, \quad   T^\dagger Y T = Y/ \lambda .
\end{equation}
Let us show that it is an incoherent transformation since it transforms eigenstates of $X$ into eigenstates of $X$, namely
\begin{equation}
\label{stb}
T | x \rangle = \sqrt{\lambda}| \lambda x\rangle ,
\end{equation}
as required by proper normalization of transformed statistics. Moreover, we may as well consider its action on the physical versions of $|x \rangle$ in Eqs. (\ref{chi}) and (\ref{wf}),
\begin{equation}
\label{stbp}
T |\chi_{j,\sigma} \rangle =  | \chi_{j,\lambda \sigma} \rangle , \quad 
T |\xi_{x,\sigma} \rangle  = | \xi_{\lambda x, \lambda\sigma} \rangle  .
\end{equation}
We see that squeezing preserve their incoherent limit $\mathcal{C}\rightarrow 0$ when $\sigma \rightarrow 0$, confirming squeezing as an incoherent transformation.

\bigskip

Next, we study the effect of the quadrature-squeezing transformation (\ref{qst}) in the coherence of any state, partially coherent in general. The coherence for the transformed state, $ T \rho T^\dagger$, becomes
\begin{equation}
  \mathcal{C} (T \rho T^\dagger) =  \int_{-\infty}^\infty\int_{-\infty}^\infty dx dx^\prime | \langle x |T \rho T^\dagger | x^\prime \rangle |.
\end{equation}
 which, taking into account that $T^\dagger=T^{-1}$, is equivalent to
\begin{equation}
  \mathcal{C} (T \rho T^\dagger) =  \frac{1}{\lambda} \int_{-\infty}^\infty\int_{-\infty}^\infty dx dx^\prime | \langle x/\lambda |\rho | x^\prime /\lambda \rangle |.
\end{equation}
After a suitable change of integration variables, this becomes
\begin{equation}
\label{ic}
\mathcal{C} (T \rho T^\dagger) =  \lambda\int_{-\infty}^\infty\int_{-\infty}^\infty dx dx^\prime | \langle x | \rho | x^\prime \rangle | = \lambda \mathcal{C} (\rho) .
\end{equation}
As a truly incoherent transformation, if the original state is incoherent, $\mathcal{C} (\rho)=0$, the transformed state is also incoherent, $\mathcal{C} (T \rho T^\dagger) = 0$ for all $\lambda$. The very same relation in Eq. (\ref{ic}) implies that squeezing improves coherence whenever $\lambda > 1$ when $\mathcal{C} \neq 0$.

\bigskip

Intuitively there seems to be no obstacle to the fact that a transformation that adds no coherence to incoherent states might add coherence to partially coherent states. Amplification and squeezing seem to be good candidates to display this behavior. It must be pointed out also that this effect is not possible in discrete bases since in such a case unitary squeezing transformations of the reference basis are not allowed, or otherwise, it leads to nonbijective canonical transformations and rather sophisticated objects such as ambiguity spin \cite{KMS78,LS91}.

In the following, we analyze whether this effect is also reproduced by the relative entropy in Eq. (\ref{re}). To this end, we note that the first factor is invariant under unitary transformations, so we focus on the behavior of the second factor with squeezing, this is $\mathrm{tr} (\rho^\prime \ln \rho^\prime_d )$ with $\rho^\prime = T \rho T^\dagger$.  By using the same family of incoherent states $|\chi_{j,\sigma}\rangle$ after the transformation,
\begin{equation}
\label{rdp2}
\rho^\prime_{\rm d} =  \sum_{j=-\infty}^\infty  \langle\chi_{j,\sigma} | T \rho T^\dagger| \chi_{j,\sigma}\rangle |\chi_{j,\sigma}\rangle \langle \chi_{j,\sigma} | ,
\end{equation}
this term becomes, 
\begin{equation}
   \mathrm{tr} \left ( \rho^\prime \ln \rho^\prime_d \right ) = \sum_{j=-\infty}^\infty \langle\chi_{j,\sigma} | T \rho T^\dagger |\chi_{j,\sigma} \rangle \ln  \langle\chi_{j,\sigma} | T \rho T^\dagger  |\chi_{j,\sigma} \rangle .
\end{equation}
After Eq. (\ref{stbp}) we can see how
\begin{equation}
     \langle\chi_{j,\sigma} | T \rho T^\dagger |\chi_{j,\sigma} \rangle = \langle\chi_{j, \sigma/\lambda } | \rho |\chi_{j, \sigma/\lambda} \rangle \simeq \frac{\sigma}{\lambda} p(j\sigma/\lambda) ,
\end{equation}
recalling that $p(x)$ is the quadrature distribution $p(x) = \langle x |\rho |x\rangle$. Then we have
\begin{equation}
   \mathrm{tr} \left ( \rho^\prime \ln \rho^\prime_d \right ) \simeq  \sum_{j=-\infty}^\infty \frac{\sigma}{\lambda}p(j\sigma/\lambda ) \ln \left [ \frac{\sigma}{\lambda} p(j\sigma /\lambda  ) \right ],
\end{equation}
that in the limit $\sigma \rightarrow 0$ becomes
\begin{equation}
   \mathrm{tr} \left ( \rho^\prime \ln \rho_d^\prime \right ) \simeq  \ln \frac{\sigma}{\lambda}  + \int_{-\infty}^\infty p(x) \ln p(x) .
\end{equation}
Finally, the coherence becomes
\begin{equation}
    S(\rho^\prime||\rho^\prime_d) =  S(\rho||\rho_d) + \ln \lambda .
\end{equation}
Therefore, this measure shows the same behaviour than measure (\ref{mc}), and squeezing increases coherence for $\lambda > 1$. Leaving aside technical details, the result is intuitive given the relation between quadrature-coherence and the quadrature uncertainty revealed by the examples considered above. Finally, we expect the same result may be obtained with pre-incoherent states $|\xi_{x,\sigma} \rangle$ and Eq. (\ref{rd}) since in the $\sigma \rightarrow 0$ they tend to be effectively orthogonal.

\bigskip

\section{Multi-mode case}

In this section, we examine the multi-mode case, starting with a pair of comments about incoherent states and the contribution of the vacuum modes. Then we consider the effect on coherence of some useful two-mode transformations such as beam splitting and two-mode squeezing. 

\bigskip

First, we note that, for factorized states, the generalization of Eq. (\ref{mc}) leads to the product of the corresponding single-mode coherence, 
\begin{equation}
\label{fact}
\rho = \rho_1 \otimes\rho_2 \ldots\otimes\rho_N, \hspace{2mm} \mathcal{C} (\rho)= \mathcal{C} (\rho_1)\mathcal{C} (\rho_2)\ldots\mathcal{C} (\rho_N)  . 
\end{equation}
As a consequence, if any of the modes, say $\rho_j$, is in an incoherent state, or better said, tends to be incoherent in the sense that $\mathcal{C} (\rho_j) \rightarrow 0$, then the whole system tends to be incoherent $\mathcal{C} (\rho) \rightarrow 0$. This does not occur regarding coherence in other bases. 

\bigskip

There is also another intriguing consequence of factorization in Eq. (\ref{fact}). In the conventional analysis of coherence, just the modes that are not always in the vacuum state $\rho_{v}$ are considered, since the vacuum modes usually do not contribute to coherence. This is not the case in the quadrature basis since the vacuum is just another Gaussian state with $\Delta X = 1/2$, so these modes also contribute to the total coherence, $\mathcal{C} (\rho_{v}) = \sqrt{2 \pi}$. So, in every situation, to compute the total amount of coherence we should consider that any $\rho$, single-mode or multi-mode, should be completed with the total number of modes that remain in the vacuum state throughout the whole process. This implies that coherence diverges in  the limit when the total number of modes tends to infinity,  $\mathcal{C} (\rho) \rightarrow \infty$.

\bigskip

\subsection{Coherence invariance under lossless beam splitting}

Beam splitting is a well-known method of producing coherence in classical optics, and so a key ingredient in most celebrated interferometers. In a recent work, we have examined its performance in the quantum-optical domain, regarding quantum coherence in the photon-number basis \cite{AL23}. Here, we examine this very same question on the quadrature basis. 

\bigskip

The action of a lossless beam splitter is usually expressed in terms of complex-amplitude operators, say 
\begin{equation}
    V^\dagger a_1 V  = t_1 a_1+r_2 a_2, \quad  V^\dagger a_2 V = r_1 a_1+t_2 a_2,
\end{equation}
where $t_{1,2}$ $r_{1,2}$ are the corresponding complex transmission and reflection coefficients  \cite{LS95}. We have already studied the effect on coherence of phase shifting, this is free propagation. So, for definiteness, we shall focus here just on the effect of beam-splitting modulus phase shifts. This is tantamount to considering real transmission and reflection coefficients, so that the beam splitter just mixes the quadratures that define the coherence basis, say 
\begin{eqnarray}
   &  V^\dagger X_1 V = \cos \theta X_1+ \sin \theta X_2 , & \nonumber \\
   & & \\
   & V^\dagger X_2 V = - \sin \theta X_1 + \cos \theta X_2, & \nonumber
\end{eqnarray}
and similarly for the $Y$ quadratures replacing $X_{1,2}$ by $Y_{1,2}$. Therefore, the action of the beam splitter on the quadrature basis of eigenvectors of $X_1$ and $X_2$ is 
\begin{equation}
\label{ttb}
V| x_1 \rangle_1 | x_2 \rangle_2 = |\cos \theta x_1 + \sin \theta x_2  \rangle_1 |-\sin \theta x_1 + \cos \theta x_2 \rangle_2 ,
\end{equation}
where $|x\rangle_j$ are the eigenstates of the quadrature $X_j$ in the corresponding mode.

\bigskip

With this transformation law, it is clear that there is no effect of beam splitting on quadrature coherence and the states $\rho$ and $V \rho V^\dagger$ have the same coherence $\mathcal{C}$. This is because the effect of the transformation on the quadrature basis (\ref{ttb}) can be compensated by a change of integration variables in Eq. (\ref{mc}) with unit Jacobian, so that 
\begin{equation}
\mathcal{C} \left ( T \rho T^\dagger \right ) = \mathcal{C} \left ( \rho  \right ) .
\end{equation}
Therefore, beam splitting is essentially an incoherent transformation regarding quadrature coherence, in sharp contrast to its behavior regarding coherence in the photon number basis \cite{AL23}.

\bigskip

\subsection{Coherence invariance under two-mode squeezing}

The same result of coherence invariance holds under two-mode squeezing
\begin{equation}
    T^\dagger a_1 T = \mu a_1 + \nu a^\dagger_2, \quad  T^\dagger a_2 T = \mu a_2 + \nu a^\dagger_1 .
\end{equation}
As before, we consider modulus phase shifts so that coefficients $\mu$ and $\nu$ are real and then
\begin{eqnarray}
   &  T X_1 T^\dagger = \cosh \theta X_1+ \sinh \theta X_2 , & \nonumber \\
   & & \\
   &T X_2 T^\dagger = \sinh \theta X_1 + \cosh \theta X_2  , & \nonumber
\end{eqnarray}
and similarly for the $Y$ quadratures replacing $X_{1,2}$ by $Y_{1,2}$ and $\sinh \theta$ by $-\sinh \theta$.  Therefore, the action on the quadrature basis of eigenvectors of $X_1$ and $X_2$ is 
\begin{equation}
T| x_1 \rangle_1 | x_2 \rangle_2 = |\cosh \theta x_1 + \sinh \theta x_2  \rangle_1 |\sinh \theta x_1 +\cosh \theta x_2  \rangle_2 ,
\end{equation}
and again there is no effect on quadrature coherence in Eq. (\ref{mc}) by compensation via a unit-Jacobian change of integration variables.

\bigskip

\section{Conclusions}

In this work, we have translated to quantum-optical quadratures an standard resource-theoretical approach to quantum coherence. We have studied the associated incoherent states and incoherent operations. These questions arise specially blurred due to the lack of physical incoherent states. We develop different approaches to define this states. As results, we found a relation of incoherence with field energy, a kind of infinite background of coherence present in the vacuum modes, and the fact that basic quantum-optical transformations, such as squeezing and beam splitting, are incoherent transformations. Moreover, we have found that squeezing may add coherence to partially coherent states, in spite of its incoherent character.

These results may serve for a better understanding of the different approaches to quantum coherence and their relation to already existing theories and concepts about coherence derived from classical and quantum optics. To this end, quantum-field quadrature may provide an interesting arena to help in the progress of this subject.

\bigskip

\section*{Acknowledgments }
L. A. and A. L. acknowledge financial support from project PR44/21--29926 from Santander Bank and Universidad Complutense of Madrid.


\begin{thebibliography}{00}

\bibitem{SLB05}
S. L. Braunstein and P. van Loock,
Quantum information with continuous variables,
\href{https://doi.org/10.1103/RevModPhys.77.513}{Rev. Mod. Phys. \textbf{77}, 513 (2005)}.

\bibitem{GF23}
G. Ferrari, L. Lami, T. Theurer, and M. B. Plenio, Asymptotic State Transformations of Continuous Variable 
Resources, \href{https://doi.org/10.1007/s00220-022-04523-6}{Commun. Math. Phys. \textbf{398}, 291--351 (2023)}.

\bibitem{ZSLF16}
Y.-R. Zhang, L.-H. Shao, Y. Li, and H. Fan, Quantifying coherence in infinite-dimensional systems, \href{http://dx.doi.org/10.1103/PhysRevA.93.012334}{Phys. Rev. A {\bf 93}, 012334 (2016)}.

\bibitem{GA18}
L. Lami, B. Regula, X. Wang, R. Nichols, A. Winter, and G. Adesso,
Gaussian quantum resource theories,
\href{https://doi.org/10.1103/PhysRevA.98.022335}{Phys. Rev. A \textbf{98}, 022335 (2018)}.

\bibitem{BY18}
B. Yadin, F. C. Binder, J. Thompson, V. Narasimhachar, M. Gu, and M.S. Kim,
Operational Resource Theory of Continuous-Variable Nonclassicality,
\href{https://doi.org/10.1103/PhysRevX.8.041038}{Phys. Rev. X \textbf{8}, 041038 (2018)}.

\bibitem{MI16}
M. Idel, D. L. and M. Wolf,
An operational measure for squeezing,
\href{http://dx.doi.org/10.1088/1751-8113/49/44/445304}{J. Phys. A: Math. Theor. \textbf{49,} 445304 (2016)} .

\bibitem{BR21}
B. Regula, L. Lami, Gi. Ferrari, and R. Takagi,
Operational Quantification of Continuous-Variable Quantum Resources,
\href{https://doi.org/10.1103/PhysRevLett.126.110403}{Phys. Rev. Lett. \textbf{126}, 110403 (2021)}.

\bibitem{LL21}
L. Lami, B. Regula, R. Takagi, and G. Ferrari,
Framework for resource quantification in infinite-dimensional general probabilistic theories,
\href{https://doi.org/10.1103/PhysRevA.103.032424}{Phys. Rev. A \textbf{103}, 032424 (2021)}.

\bibitem{HKPU21}
E. Haapasalo, T. Kraft, J.-P. Pellonp\"a\"a, and R. Uola,
Operational Characterization of Infinite-Dimensional Quantum Resources,
\href{https://doi.org/10.1103/PhysRevLett.126.110403}{Phys. Rev. Lett. {\bf 127}, 250401 (2021)}.

\bibitem{AL22}
L. Ares and A. Luis, Distance-based approach to quantum coherence and nonclassicality, 
\href{https://doi.org/10.1103/PhysRevA.106.012415}{Phys. Rev. A {\bf 106}, 012415 (2022)}.

\bibitem{LGT17}
C. L. Liu, Y.-Q. Guo,2 and D. M. Tong,
Enhancing coherence of a state by stochastic strictly incoherent operations,
\href{https://doi.org/10.1103/PhysRevA.96.062325}{Phys. Rev. A {\bf 96}, 062325 (2017)}.

\bibitem{LZ20}
C. L. Liu  and D. L. Zhou,
Increasing the dimension of the maximal pure coherent subspace of a state via incoherent operations,
\href{https://doi.org/10.1103/PhysRevA.102.062427}{Phys. Rev. A {\bf 102}, 062427 (2020)}.

\bibitem{BCP14}
T. Baumgratz, M. Cramer, and M. B. Plenio, Quantifying Coherence, \href{https://doi.org/10.1103/PhysRevLett.113.140401}{Phys. Rev. Lett. {\bf 113}, 140401 (2014)}.

\bibitem{HB20}
A. Hertz and S. De Bi\` evre,
Quadrature Coherence Scale Driven Fast Decoherence of Bosonic Quantum Field States
\href{https://doi.org/10.1103/PhysRevLett.124.090402}{Phys. Rev. Lett. {\bf 124}, 090402 (2020)}.

\bibitem{GTH23}
A. Z. Goldberg, G. S. Thekkadath, and K. Heshami,
Measuring the quadrature coherence scale on a cloud quantum computer,
\href{https://doi.org/10.1103/PhysRevA.107.042610}{Phys. Rev. A {\bf 107}, 042610 (2023)}.

\bibitem{CT06}
T. M. Cover and J. A. Thomas, {\it Elements of Information Theory. Second Edition.} (Wiley Interscience 2006).

\bibitem{KMS78}
P. Kramer, M. Moshinsky, and T. H. Seligman,
Nonbijective canonical transformations and their representation in quantum mechanics,
\href{https://doi.org/10.1063/1.523712}{J. Math. Phys. {\bf 19}, 683--693 (1978)}.

\bibitem{LS91}
A. Luis and L. L. S\'anchez-Soto, Nonclassical states of light and canonical transformations,
\href{https://doi.org/10.1088/0305-4470/24/9/018}{J. Phys. A: Math. Gen. \textbf{24}, 2083--2092 (1991)}.

\bibitem{AL23}
L. Ares and A. Luis, Beam splitter as quantum coherence-maker, \href{https://doi.org/10.1088/1402-4896/aca1e7}{Physica Scripta, {\bf 98}, 015101 (2023)}.

\bibitem{LS95}
A. Luis and L. L. S\' anchez-Soto, A quantum description of the beam splitter, \href{https://doi.org/10.1088/1355-5111/7/2/005}{Quantum Semiclass. Opt. {\bf 7}, 153--160  (1995)}. 

\end{thebibliography}
\end{document}